\documentclass[conference]{IEEEtran}
\IEEEoverridecommandlockouts
\usepackage{cite}
\usepackage{amsmath,amssymb,amsfonts}
\usepackage{algorithmic}
\usepackage{graphicx}
\usepackage{textcomp}
\usepackage{xcolor}

\usepackage{color}
\usepackage{makecell}
\usepackage{multirow}
\usepackage{tabularx}
\usepackage{graphicx}
\usepackage{amsmath}
\usepackage{booktabs}
\def\tablename{Table}
\usepackage{caption}

\def\BibTeX{{\rm B\kern-.05em{\sc i\kern-.025em b}\kern-.08em
    T\kern-.1667em\lower.7ex\hbox{E}\kern-.125emX}}
\begin{document}

\title{Model-Guardian: Protecting against Data-Free Model Stealing Using Gradient Representations and Deceptive Predictions$^{*}$
\thanks{$^{*}$This is the full version of the paper accepted by ICME 2025. This work is supported by Beijing Municipal Science \& Technology Commission: New Generation of Information and Communication Technology Innovation - Research and Demonstration Application of Key Technologies for Privacy Protection of Massive Data for Large Model Training and Application (Z231100005923047).}
\thanks{$^{\dagger}$Corresponding author.}
}

\author{
    \IEEEauthorblockN{Yunfei Yang$^{1,2,3}$, Xiaojun Chen$^{1,2,3,\dagger}$, Yuexin Xuan$^{1,2,3}$, and Zhendong Zhao$^{1,2}$}
    \IEEEauthorblockA{$^{1}$Institute of Information Engineering, Chinese Academy of Sciences, Beijing, China}
    \IEEEauthorblockA{$^{2}$State Key Laboratory of Cyberspace Security Defense, Beijing, China}
    \IEEEauthorblockA{$^{3}$School of Cyber Security, University of Chinese Academy of Sciences, Beijing, China}
    \IEEEauthorblockA{\{yangyunfei, chenxiaojun, xuanyuexin, zhaozhendong\}@iie.ac.cn}
}

\maketitle

\begin{abstract}
Model stealing attack is increasingly threatening the confidentiality of machine learning models deployed in the cloud. Recent studies reveal that adversaries can exploit data synthesis techniques to steal machine learning models even in scenarios devoid of real data, leading to data-free model stealing attacks. Existing defenses against such attacks suffer from limitations, including poor effectiveness, insufficient generalization ability, and low comprehensiveness. In response, this paper introduces a novel defense framework named Model-Guardian. Comprising two components, Data-Free Model Stealing Detector (DFMS-Detector) and Deceptive Predictions (DPreds), Model-Guardian is designed to address the shortcomings of current defenses with the help of the artifact properties of synthetic samples and gradient representations of samples. Extensive experiments on seven prevalent data-free model stealing attacks showcase the effectiveness and superior generalization ability of Model-Guardian, outperforming eleven defense methods and establishing a new state-of-the-art performance. Notably, this work pioneers the utilization of various GANs and diffusion models for generating highly realistic query samples in attacks, with Model-Guardian demonstrating accurate detection capabilities.
\end{abstract}

\begin{IEEEkeywords}
AI Security, Model Stealing, Defense
\end{IEEEkeywords}

\section{Introduction}
\vspace{-0.1cm}
With the rapid growth of deep learning, some companies offer Machine Learning as a Service (MLaaS), deploying pre-trained models in the cloud for commercial profit. However, vulnerabilities in MLaaS have been exposed, particularly regarding model stealing attacks~\cite{truong2021data,Rosenthal2023DisGUIDEDD,yang2024stms}. In these attacks, adversaries query APIs with crafted samples to acquire annotated data for training equivalent models~\cite{sanyal2022towards}. Attackers can then use the stolen model for adversarial attacks or membership inference attacks. Recent studies show that even without real data or knowledge of the original distribution for training data, attackers can still achieve high  attack performance using data-free stealing methods~\cite{kariyappa2021maze,beetham2022dual,lin2023quda,yang2024dualcos}, a trend that has gained significant attention in the research community.

To mitigate the risk of model stealing attacks, various defenses have been proposed, categorized as active and passive. Active defenses~\cite{kariyappa2020defending,cheng2023apgp,guo2023isolation} perturb outputs during the prediction phase to hinder attackers from replicating the victim model’s functionality. Passive defenses~\cite{juuti2019prada,yao2023fdinet} detect anomalies in query sequences, leading to service termination. However, current defenses have limitations against data-free stealing attacks: (1) Most lack data-free attack considerations, as shown in our experiments, making them ineffective and limited in generalization. (2) Active defenses cannot distinguish and terminate malicious queries, allowing attackers to generate clone models with acceptable performance by inflating queries, while reducing benign query accuracy. (3) Passive defenses, which rely on query distribution characteristics, fail to detect individual samples and could improve in detection accuracy and false positive rates (FPR).

To address these limitations, we propose Model-Guardian, a comprehensive defense safeguarding model privacy from data-free model stealing attacks. It consists of Data-Free Model Stealing Detector (DFMS-Detector) and Deceptive Predictions (DPreds) modules. Due to the varied generative model structures and diverse generated samples in data-free model stealing, we adopt more generalized representations, gradients, as training data for DFMS-Detector. In the DPreds module, we strategically perturb model outputs to maintain class probability relationships while providing users with perturbed results. During implementation, Model-Guardian adapts its responses based on query detection results.
The proposed Model-Guardian exhibits several advantages: (1) Effectiveness: It can effectively defend against prevalent data-free stealing attacks. (2) Generalization: It performs well against novel attacks and data from various generative models. (3) Harmlessness: It minimally impacts benign users, preserving model accuracy.

\noindent\textbf{Contributions.} Our contributions encompass four aspects:
\begin{itemize}
    \item Proposal of Model-Guardian, an innovative and effective defense addressing data-free model stealing attacks. This work represents a pioneering effort, specifically targeting data-free scenarios and considering adversaries' use of GANs and diffusion models for realistic data synthesis.
    \item Enhancement of generalization through the use of gradient representations in training an ensemble detector, DFMS-Detector, within Model-Guardian. This detector accurately discerns malicious queries from multiple attacks, demonstrating low FPR on benign samples.
    \item Introduction of a simple yet effective prediction perturbation algorithm, Deceptive Predictions (DPreds), capable of disrupting the probability distribution and adversary's clone training process without altering the magnitude relationship of original class probabilities. Importantly, it does not impact the use of misjudged benign users.
    \item Conduct of extensive experiments across multiple datasets to evaluate the effectiveness of Model-Guardian. Results showcase superior performance compared to baselines in defending against data-free stealing attacks.
\end{itemize}

\section{Related Work}
\vspace{-0.1cm}
\subsection{Model Stealing Attacks}

\vspace{-0.1cm}
\textbf{Adversarial Generation-based Attacks.}
The attacker, possessing limited training data, increases query data by synthesizing adversarial samples.
For instance, JBDA~\cite{papernot2017practical} introduces a data augmentation technique based on Jacobian matrix.

\textbf{Proxy Data-based Attacks.}
Attackers are constrained to utilizing public datasets as query samples, with an emphasis on optimizing sampling strategies to enhance stealing effectiveness. KnockoffNet~\cite{orekondy2019knockoff} leverages reinforcement learning to construct a transfer set.
MExMI~\cite{xiao2022mexmi} represents progress by concurrently incorporating model stealing and membership inference attacks, achieving mutual performance enhancement.

\textbf{Data-Free Attacks.}
The attacker, lacking access to real data, relies on synthesizing samples through a generator. DaST~\cite{zhou2020dast} employs a multi-branch architecture and label-controlled loss for generator to synthesize data. MAZE~\cite{kariyappa2021maze} and DFME~\cite{truong2021data} utilize zero-order gradient estimation to compute the victim's gradient, facilitating training update of generator. DFMS~\cite{sanyal2022towards} conducts data-free stealing attacks under hard-label settings. DisGUIDE~\cite{Rosenthal2023DisGUIDEDD} and Dual Student~\cite{beetham2022dual} are pioneering in using the dual clone structures for stealing. QUDA~\cite{lin2023quda} introduces a novel attack leveraging a frozen GAN pre-trained with publicly irrelevant data to provide weak image priors. 


Data-free model stealing attacks are more practical in real-world scenarios due to challenges in obtaining victim training data and the gap between public datasets and victim's private training sets. Therefore, our research focuses on developing defenses against data-free model stealing.


\vspace{-0.1cm}
\subsection{Defenses against Model Stealing Attacks}

\vspace{-0.1cm}
\textbf{Active Defenses.}
Defenders add perturbations or introduce randomness to model outputs, diminishing the accuracy of clone training. RS~\cite{lee2019defending} introduces fuzziness through shrewd noise addition to output probabilities. MAD~\cite{orekondy2020prediction} disrupts adversary-obtained gradients by adding perturbations to the original output. AM~\cite{kariyappa2020defending} detects OOD input presence, returning a misinformative prediction to adversary. EDM~\cite{kariyappa2021protecting} introduces randomness into the model output via an ensemble of diverse models. NT~\cite{ma2021undistillable} utilizes a Nasty Teacher to prevent model distillation and stealing. PoW~\cite{dziedzic2022increasing} significantly increases adversary's query costs based on proof of work. APGP~\cite{cheng2023apgp} ensures complete accuracy preservation for black-box model privacy. InI~\cite{guo2023isolation} directly trains defensive models by isolating the adversary's training gradient.

\textbf{Passive Defenses.}
Defenders discern model stealing attacks by identifying abnormal behavior in continuous query samples, followed by implementing subsequent defense measures. PRADA~\cite{juuti2019prada} detects malicious queries by assessing the statistical distribution of the minimum $l_{2}$-norm distance between queries. SEAT~\cite{zhang2021seat} employs a Similarity Encoder to encode visually similar samples with similar inter-code distances, determining malicious behavior based on the pair number of similar samples. FDINET~\cite{yao2023fdinet} introduces Feature Distortion Index (FDI), enabling defenders to train a binary detector for detecting the presence of stealing attacks using FDI.


Existing active defenses struggle with impacting benign queries and only reduce clone model performance without preventing attacks. Passive defenses suffer from low detection accuracy, high false positives, and susceptibility to bypassing via discontinuous malicious queries. Moreover, there is no dedicated defense for data-free model stealing, which is the primary focus of our work.

\section{Defense Strategy: Model-Guardian}

\begin{figure*}[t]
\centering
\includegraphics[width=0.85\textwidth]{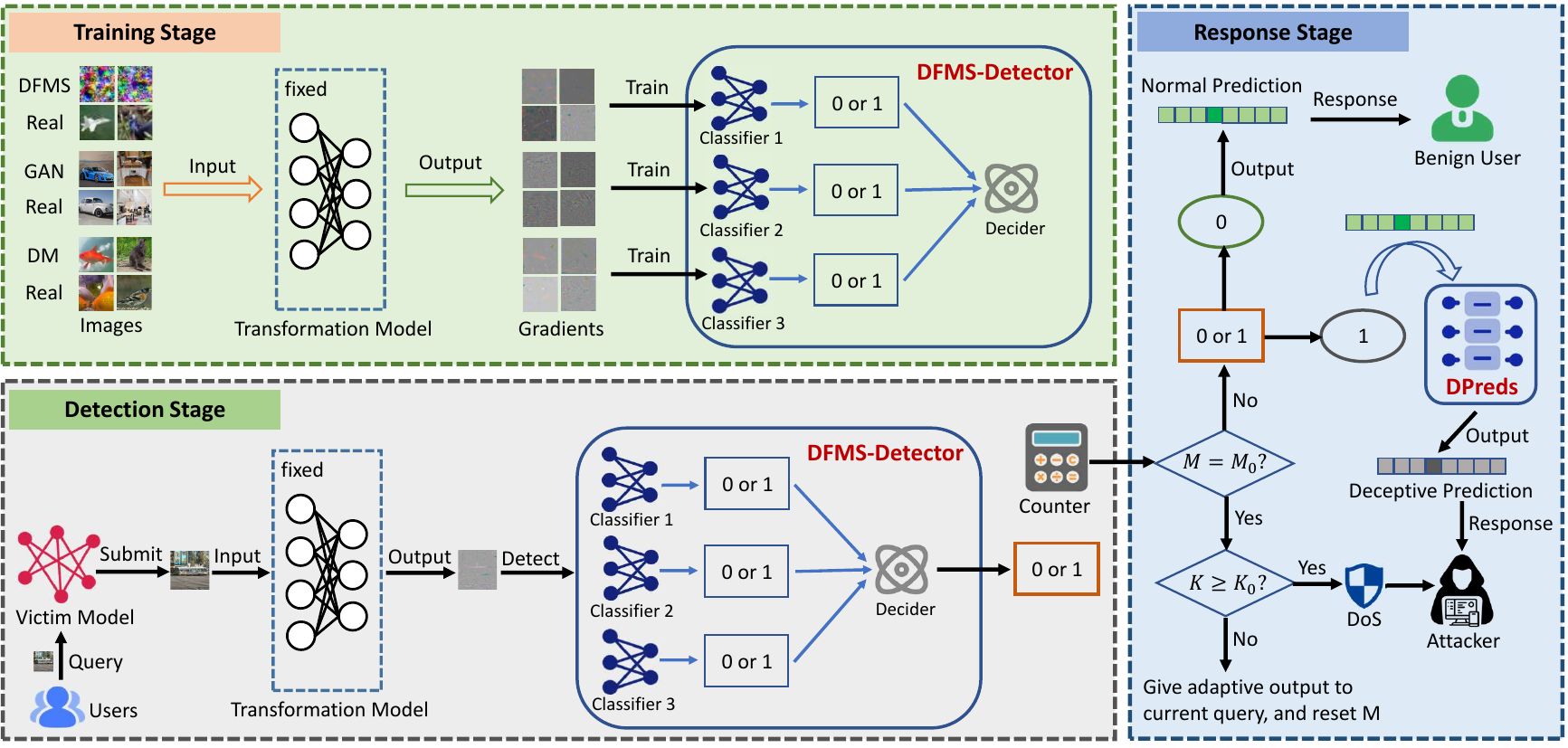}
\vspace{-0.1cm}
\caption{\textbf{Overview of our proposed Model-Guardian.} During training, synthetic images from a randomly selected Data-Free Model Stealing (DFMS) attack, Generative Adversarial Network (GAN), and Diffusion Model (DM) are combined with real images to form three distinct training sets. A pre-trained transformation model converts these data into gradients, which are used to train three sub-detectors, later integrated into a unified detector. In the detection phase, query samples of user are converted into gradients and passed to the detector for evaluation. During the response phase, the model adapts its output based on the query record.}
\label{fig2}
\vspace{-0.3cm}
\end{figure*}

\vspace{-0.1cm}
\subsection{Threat Model}
\vspace{-0.1cm}
\textbf{Adversary’s Goals and Capabilities.}
Our work focuses on model functionality stealing in image classification, wherein adversary aims to acquire a clone model $C$ imitating the functionality of victim model $V$. Specifically, adversary seeks to maximize classification accuracy $Acc(C(x; \theta_{C}), y)$ on the victim model's test set $D_{V}$, formalized as:
\begin{equation}
\setlength\abovedisplayskip{2.5pt}
\setlength\belowdisplayskip{2.5pt}
\mathop{\mathrm{argmax}}\limits_{\theta_{C}}E_{(x, y) \sim D_{V}}[Acc(C(x; \theta_{C}), y)],
\end{equation}
\noindent where $\theta_{C}$ represents the parameter of clone model $C$, and $y$ denotes the ground truth label of test sample $x$.

In practical scenarios, adversaries have limited knowledge of the victim model's architecture, parameters, hyperparameters, and training/testing sets, and can only interact through query APIs. This work adopts a soft-label setting, where the victim model outputs softmax probabilities for user-friendliness. In more realistic settings, adversaries must rely on generative models to synthesize query samples, as they cannot access real samples resembling the defender's private training data. Thus, the clone model $C$'s objective is to minimize the prediction difference between itself and the victim model $V$ on the synthetic query dataset $D_{Q}$:
\begin{equation}
\setlength\abovedisplayskip{2.5pt}
\setlength\belowdisplayskip{2.5pt}
\mathop{\mathrm{argmin}}\limits_{\theta_{C}}E_{x \sim D_{Q}}[d(C(x; \theta_{C}), V(x; \theta_{V}))],
\end{equation}
\noindent where $d(\cdot, \cdot)$ is an indicator measuring difference distance between two objects, and $\theta_{V}$ is parameters of victim model.


\textbf{Defender’s Goals and Capabilities.}
Defenders lack precise knowledge of the attacker's attack method and duration, but they can use various methods such as anomaly detection and prediction perturbation to resist attacks. The defense objective is twofold: firstly, to impede the adversary's attack, resulting in a clone model with diminished accuracy, and secondly, to minimize interference with the normal use of benign users. This goal can be formalized as follows:
\begin{equation}
\setlength\abovedisplayskip{2.5pt}
\setlength\belowdisplayskip{2.5pt}
\begin{aligned}
\mathop{\mathrm{argmin}}\limits_{\theta_{V}}&E_{(x, y) \sim D_{V}}[Acc(C(x; \theta_{C}), y)],\\
\text{s.t.}~&E_{(x, y) \sim D_{V}}[Acc(V(x; \theta_{V}), y)]\geq T,
\end{aligned}
\end{equation}
\noindent where $T$ represents the acceptable threshold for the accuracy of the victim model after degradation.


\vspace{-0.1cm}
\subsection{Overview of Model-Guardian}
\vspace{-0.1cm}
An overview of our method is shown in Figure \ref{fig2}. Generated images from three sources, along with real images, are converted into gradients. Each sub-detector is trained separately, and their outputs are integrated into an ensemble detector. For each query, the sub-detectors evaluate its maliciousness; if any sub-detector flags the sample as malicious, DFMS-Detector labels it as such. Based on the detection results, benign samples receive normal predictions, while malicious ones trigger the DPreds module for deceptive predictions. Our framework tracks the total queries and malicious samples, and if the proportion of malicious samples in $M_{0}$ queries exceeds $K_{0}$, the user's subsequent queries are terminated.

\vspace{-0.1cm}
\subsection{Training DFMS-Detector}
\label{sec4.2}
\vspace{-0.1cm}
\textbf{Collecting Training Data.}
To improve defense generalization and address common sample synthesis methods, we collect a limited training dataset from three generative sources, each representing a fake class. These sources include data from general data-free model stealing, GANs, and diffusion models. For each synthetic data type, an equal amount of real data is selected as the real class. We focus on the data synthesized by GAN and diffusion model due to attackers' ability, with advancing image generation technology, to create realistic fake images that closely resemble the victim model's test distribution, potentially bypassing conventional defenses.

\begin{figure}[t]
\centering
\includegraphics[width=0.4\textwidth]{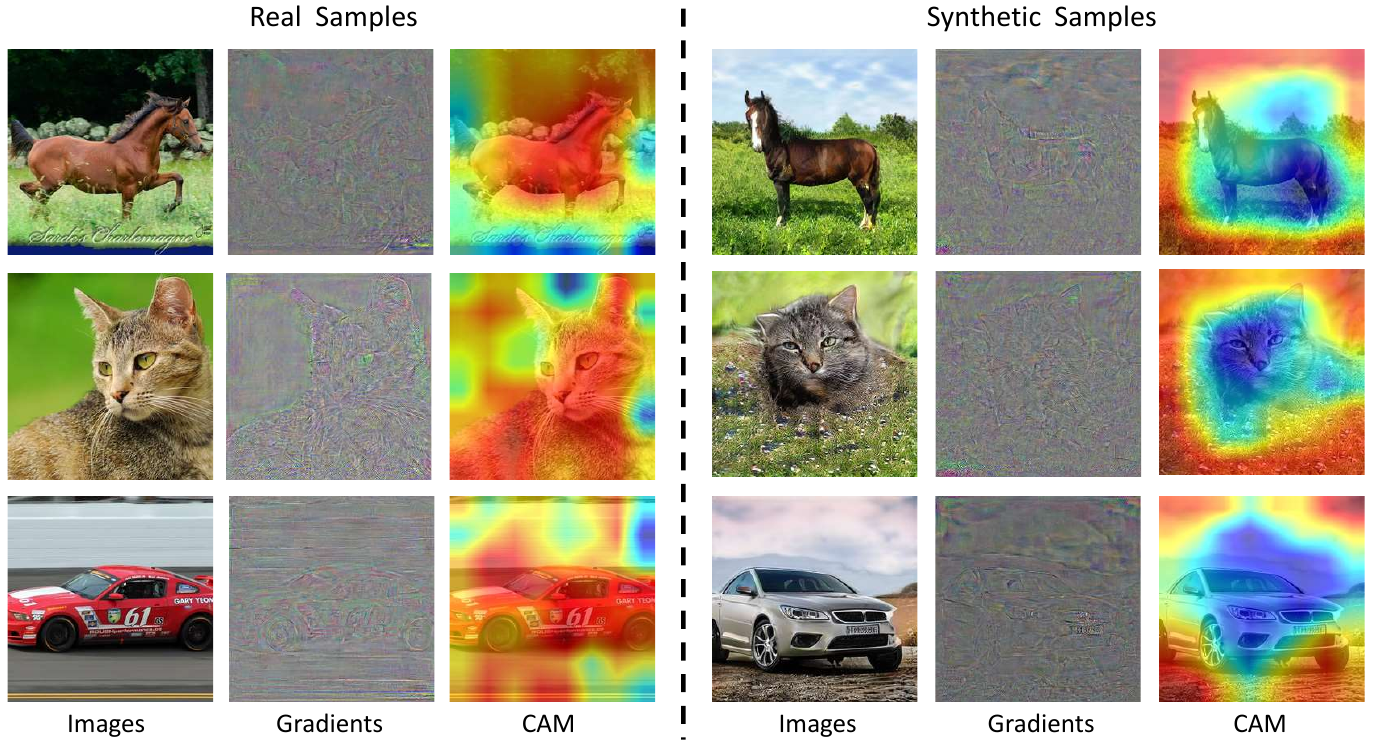}
\vspace{-0.1cm}
\caption{Visualization of gradients and Class Activate Map (CAM) extracted from detector on real and synthetic images.}
\label{fig3}
\vspace{-0.3cm}
\end{figure}

\textbf{Transforming Images to Gradients.}
Our theoretical analysis in Appendix-B and experimental results in Table \ref{tab:Contribution of Different Modules} of Appendix-E show that directly training the detector with image-form training data has limited effectiveness and generalization. In line with insights from prior works \cite{wang2020cnn,li2022defending,liu2023detection,tan2023learning}, we use gradients, a more generalized representation, as the final training data. Gradients filter out image content, retaining only key pixels. Figure \ref{fig3} illustrates the transformation from image data to gradients, with Class Activation Maps (CAM) extracted from our detector. A pre-trained CNN model is used to convert image data into gradient data, forming a novel training set.

Given a training dataset $D_{I}=\{(x_{i}, y_{i})\}_{i=1}^{n}$, where $y_{i}$ represents the label of $x_{i}$ with two classes: real ($y=0$) and fake ($y=1$). We introduce the transformation model $M_{T}(\cdot)$. The initial step involves inputting $x_{i}$ into $M_{T}$ to obtain an output feature vector $u$:
\begin{equation}
\setlength\abovedisplayskip{2.5pt}
\setlength\belowdisplayskip{2.5pt}
u=M_{T}(x_{i}).
\end{equation}

Subsequently, we compute the gradient of $sum(u)$ with respect to input $x_{i}$:
\begin{equation}
\setlength\abovedisplayskip{2.5pt}
\setlength\belowdisplayskip{2.5pt}
g=\frac{\partial sum(u)}{\partial x_{i}},
\end{equation}
\noindent where $sum(\cdot)$ is a summation operation, and $g$ serves as the generalized sample representation.

\textbf{Training the Detector.}
We normalize the gradients to the range 0–255 for training each sub-detector. Each sub-detector is a binary classifier optimized to distinguish whether an input gradient is from a fake image. The training objective for each sub-detector $F$ is as follows:
\begin{equation}
\label{eqn6}
\setlength\abovedisplayskip{2.5pt}
\setlength\belowdisplayskip{2.5pt}
L=E_{(g, y) \sim D_{G}}[CE(F(g), y)],
\end{equation}
\noindent where $CE(\cdot,\cdot)$ represents cross-entropy loss, and $D_{G}$ is the gradient dataset with labels matching the original images. After training, the sub-detectors are combined in parallel to form the ensemble DFMS-Detector.

\vspace{-0.1cm}
\subsection{Detecting Query Sample}
\vspace{-0.1cm}
\textbf{Detecting with Each Sub-Detector.}
For user queries, we transform them into gradients and pass them to DFMS-Detector. Each sub-detector independently assesses whether the sample is malicious.

\textbf{Ensembling and Making the Final Decision.}
A decision-maker evaluates the sample based on outputs of sub-detector. If at least one sub-detector flags the sample as fake ($y=1$), it is classified as malicious; otherwise, it is benign. A query counter tracks the total number of queries and those identified as malicious.

\vspace{-0.1cm}
\subsection{Returning Adaptive Output}
\vspace{-0.1cm}
For each query, output strategies depend on the detection results and the number of malicious queries. If the total queries exceed $M_{0}$ and the proportion of malicious queries exceeds $K_{0}$, further queries are terminated. Otherwise, adaptive responses are used.

\textbf{Normal Predictions for Benign Samples.}
If the decision-maker classifies a sample as normal, the original prediction (softmax probability) is returned without modification, ensuring benign users are unaffected.

\textbf{Deceptive Predictions for Malicious Samples.}
If a query is identified as malicious, the DPreds module is activated to provide deceptive predictions. DPreds is an effective active defense method proposed by us. Adversaries need unaltered probabilities to train clone models, while benign users only require relative class probabilities. Our perturbation algorithm influences malicious queries’ outputs while evading detection by adversaries and preserving the order of class labels before and after perturbation.

In the implementation, 
an identical perturbation value $r$ is added to each class probability of the original prediction $P\!=\![p_{1}, p_{2}, ..., p_{k}]$, followed by normalization to ensure a sum of one as follows:
\begin{equation}
\label{eqn8}
\setlength\abovedisplayskip{2.5pt}
\setlength\belowdisplayskip{2.5pt}
\begin{aligned}
P'&=[p_{1}+r, p_{2}+r, ..., p_{k}+r],\\
P''&=[p''_{1}, p''_{2}, ..., p''_{k}]=\frac{P'}{\sum_{i=1}^{k}(p_{i}+r)}.
\end{aligned}
\end{equation}


The modified prediction result $P''$ is returned to the malicious user, minimizing useful information and hindering effective clone model training.

\begin{table*}[ht]
    \centering
    \footnotesize
    \caption{Performance comparison (\%) with popular active defenses.}
    \label{tab:Comparison with Popular Active Defenses}
    \vspace{-0.1cm}
    \begin{tabular}{llcccccccc}
        \toprule
        \multirow{2}{*}{\textbf{Dataset}} & \multirow{2}{*}{\textbf{Method}} & \multirow{2}{*}{\textbf{BAcc.}} & \multicolumn{7}{c}{\textbf{CAcc. on Testing Attack Method}} \\
        \cmidrule(lr){4-10}
        & & & DaST & MAZE & DFME & DFMS & DisGUIDE & Dual Student & QUDA \\
        \midrule
        \multirow{10}{*}{CIFAR-10} & No Defense & 95.54 & 35.18 & 45.60 & 88.10 & 91.24 & 94.02 & 91.36 & 86.94 \\
        & RS & 91.81 (-3.73) & 26.39 & 36.42 & 67.84 & 73.90 & 69.08 & 74.41 & 62.69 \\
        & MAD & 95.16 (-0.38) & 27.44 & 35.57 & 70.53 & 75.73 & 69.57 & 79.48 & 72.16 \\
        & AM & 94.62 (-0.92) & 32.27 & 42.96 & 82.02 & 85.79 & 88.19 & 84.05 & 82.33 \\
        & EDM & 95.06 (-0.48) & 32.01 & 42.76 & 83.70 & 83.94 & 89.32 & 87.71 & 83.47 \\
        & NT & 94.72 (-0.82) & 23.22 & 30.58 & 61.79 & 71.16 & 68.87 & 63.95 & 60.86 \\
        & PoW & 93.22 (-2.32) & 32.96 & 41.83 & 81.76 & 84.22 & 87.63 & 84.96 & 80.68 \\
        & APGP & \textbf{95.54 (-0.00)} & 21.36 & 26.45 & 52.27 & 58.93 & 61.13 & 59.38 & 55.64 \\
        & InI & 94.97 (-0.57) & 31.12 & 41.45 & 78.67 & 83.12 & 86.69 & 83.05 & 76.43 \\
        & Model-Guardian & \textbf{95.54 (-0.00)} & \textbf{10.63} & \textbf{11.27} & \textbf{11.81} & \textbf{12.95} & \textbf{14.93} & \textbf{12.68} & \textbf{13.72} \\
        \midrule
        \multirow{10}{*}{CIFAR-100} & No Defense & 78.52 & 9.48 & 17.81 & 26.46 & 48.83 & 69.47 & 46.52 & 29.35 \\
        & RS & 75.53 (-2.99) & 7.58 & 13.39 & 19.53 & 38.19 & 53.74 & 33.02 & 22.89 \\
        & MAD & 76.95 (-1.57) & 7.51 & 14.25 & 21.43 & 37.60 & 58.72 & 34.87 & 23.95 \\
        & AM & 76.49 (-2.03) & 8.91 & 16.14 & 24.08 & 46.01 & 63.94 & 42.57 & 27.24 \\
        & EDM & 77.19 (-1.33) & 8.84 & 16.39 & 25.14 & 44.19 & 63.15 & 42.71 & 26.68 \\
        & NT & 77.56 (-0.96) & 7.23 & 11.97 & 18.76 & 33.84 & 50.85 & 32.33 & 21.48 \\
        & PoW & 76.71 (-1.81) & 8.66 & 16.13 & 24.24 & 45.80 & 65.72 & 41.89 & 27.39 \\
        & APGP & \textbf{78.52 (-0.00)} & 5.86 & 11.31 & 15.22 & 29.79 & 42.38 & 29.30 & 18.93 \\
        & InI & 78.05 (-0.47) & 8.22 & 16.46 & 23.07 & 44.34 & 64.54 & 40.71 & 26.83 \\
        & Model-Guardian & \textbf{78.52 (-0.00)} & \textbf{2.59} & \textbf{3.08} & \textbf{5.19} & \textbf{8.18} & \textbf{9.64} & \textbf{7.03} & \textbf{7.26} \\
        \midrule
        \multirow{10}{*}{ImageNet} & No Defense & 62.08 & 8.86 & 15.47 & 23.84 & 43.69 & 61.64 & 45.88 & 24.97 \\
        & RS & 58.87 (-3.21) & 7.06 & 11.72 & 18.69 & 31.78 & 48.29 & 32.69 & 18.58 \\
        & MAD & 60.66 (-1.42) & 7.18 & 11.81 & 19.32 & 35.01 & 47.45 & 37.85 & 19.56 \\
        & AM & 59.74 (-2.34) & 8.25 & 14.16 & 22.02 & 40.04 & 57.50 & 43.22 & 23.74 \\
        & EDM & 60.22 (-1.86) & 8.12 & 14.24 & 22.33 & 40.45 & 57.46 & 42.97 & 23.48 \\
        & NT & 61.05 (-1.03) & 6.23 & 10.79 & 16.44 & 31.62 & 43.98 & 34.26 & 18.35 \\
        & PoW & 59.44 (-2.64) & 7.96 & 14.68 & 21.86 & 41.37 & 57.68 & 43.14 & 22.94 \\
        & APGP & \textbf{62.08 (-0.00)} & 5.17 & 9.22 & 14.90 & 24.63 & 38.43 & 29.51 & 14.32 \\
        & InI & 61.39 (-0.69) & 7.77 & 13.87 & 22.08 & 38.69 & 54.03 & 42.09 & 22.85 \\
        & Model-Guardian & \textbf{62.08 (-0.00)} & \textbf{2.25} & \textbf{2.51} & \textbf{4.52} & \textbf{10.33} & \textbf{14.41} & \textbf{11.02} & \textbf{4.73} \\
        \bottomrule
    \end{tabular}
\end{table*}

\begin{table*}[t]
    \centering
    \caption{\textbf{Performance comparison (\%) with popular passive defenses.} To ensure consistency with existing research, we evaluate only the detection performance of our DFMS-Detector. Existing methods assess whether each batch is a sequence of malicious queries. Thus, the results in this table represent the detection accuracy and false positive rate of each method across all batch sequences, indicating the proportion of batch sequences that are identified as malicious.}
    \label{tab:Comparison with Popular Passive Defenses}
    \vspace{-0.1cm}
    \resizebox{\textwidth}{!}{
    \begin{tabular}{lccccccccccccccc}
        \toprule
        \multirow{2}{*}{\textbf{Method}} & \multirow{2}{*}{$bs$} & \multicolumn{2}{c}{\textbf{DaST}} & \multicolumn{2}{c}{\textbf{MAZE}} & \multicolumn{2}{c}{\textbf{DFME}} & \multicolumn{2}{c}{\textbf{DFMS}} & \multicolumn{2}{c}{\textbf{DisGUIDE}} & \multicolumn{2}{c}{\textbf{Dual Student}} & \multicolumn{2}{c}{\textbf{QUDA}} \\
        \cmidrule(lr){3-4} \cmidrule(lr){5-6} \cmidrule(lr){7-8} \cmidrule(lr){9-10} \cmidrule(lr){11-12} \cmidrule(lr){13-14} \cmidrule(lr){15-16}
        & & DAcc. & FPR & DAcc. & FPR & DAcc. & FPR & DAcc. & FPR & DAcc. & FPR & DAcc. & FPR & DAcc. & FPR \\
        \midrule
        \multirow{2}{*}{PRADA} & 50 & 0.00 & 0.00 & 0.00 & 0.00 & 0.00 & 0.00 & 0.00 & 0.00 & 0.00 & 0.00 & 0.00 & 0.00 & 0.00 & 0.00 \\
        & 500 & 18.00 & 34.00 & 22.00 & 34.00 & 32.00 & 34.00 & 13.00 & 34.00 & 25.00 & 34.00 & 21.00 & 34.00 & 16.00 & 34.00 \\
        \multirow{2}{*}{SEAT} & 50 & 13.60 & 0.80 & 10.20 & 0.90 & 11.80 & 0.80 & 9.40 & 0.70 & 13.80 & 0.80 & 12.70 & 0.80 & 11.30 & 0.90 \\
        & 500 & 72.00 & 6.00 & 65.00 & 6.00 & 76.00 & 6.00 & 54.00 & 6.00 & 81.00 & 6.00 & 77.00 & 6.00 & 73.00 & 6.00 \\
        \multirow{2}{*}{FDINET} & 50 & 86.40 & 1.80 & 89.30 & 1.70 & 89.90 & 1.80 & 89.80 & 1.60 & 91.60 & 1.70 & 92.20 & 1.80 & 94.10 & 1.90 \\
        & 500 & 95.00 & 2.00 & 98.00 & 2.00 & \textbf{100.00} & 2.00 & 97.00 & 2.00 & 99.00 & 2.00 & \textbf{100.00} & 2.00 & \textbf{100.00} & 2.00 \\
        \multirow{2}{*}{\makecell[l]{Model-\\Guardian}} & 50 & \textbf{97.30} & 0.60 & \textbf{98.10} & 0.40 & \textbf{99.60} & 0.20 & \textbf{99.20} & 0.10 & \textbf{98.70} & 0.40 & \textbf{98.50} & 0.30 & \textbf{97.90} & 0.50 \\
        & 500 & \textbf{98.00} & 2.00 & \textbf{99.00} & 2.00 & \textbf{100.00} & 0.00 & \textbf{99.00} & 0.00 & \textbf{100.00} & 2.00 & \textbf{100.00} & 1.00 & 99.00 & 1.00 \\
        \bottomrule
    \end{tabular}
    }
    \vspace{-0.3cm}
\end{table*}

\section{Experiments}

\vspace{-0.1cm}
\subsection{Experimental Setup}
\vspace{-0.1cm}
\textbf{Datasets and Architectures.}
In the defense testing phase, we perform experiments on: (1) defense against data-free stealing attacks on CIFAR-10, CIFAR-100, and ImageNet; (2) detection on datasets generated by seven GAN models, including ProGAN~\cite{karras2018progressive}, StyleGAN~\cite{karras2019style}, StyleGAN2~\cite{karras2020analyzing}, BigGAN~\cite{brock2018large}, CycleGAN~\cite{zhu2017unpaired}, StarGAN~\cite{choi2018stargan}, and GauGAN~\cite{park2019semantic}; (3) detection on datasets generated by six diffusion models, including ADM~\cite{dhariwal2021diffusion}, SD-v1~\cite{rombach2022high}, Guided~\cite{dhariwal2021diffusion}, DALL-E~\cite{ramesh2021zero}, LDM~\cite{rombach2022high}, and Glide~\cite{nichol2022glide}.

We use ResNet-34 as the victim model and ResNet-18 as the clone model. The transformation model is the StyleGAN discriminator pre-trained on LSUN-bedroom, and the classifier for all sub-detectors is a pre-trained ResNet-50 on ImageNet. Details of DFMS-Detector's training set are in Appendix-C.

\textbf{Attack Methods.}
We evaluate our defense against seven attack methods: DaST~\cite{zhou2020dast}, MAZE~\cite{kariyappa2021maze}, DFME~\cite{truong2021data}, DFMS~\cite{sanyal2022towards}, DisGUIDE~\cite{Rosenthal2023DisGUIDEDD}, Dual Student~\cite{beetham2022dual}, and QUDA~\cite{lin2023quda}. Additionally, we generate malicious query samples using seven GAN models and six diffusion models to assess detection performance.

\textbf{Defense Methods.}
We compare our method with eleven defenses: eight active defenses (RS~\cite{lee2019defending}, MAD~\cite{orekondy2020prediction}, AM~\cite{kariyappa2020defending}, EDM~\cite{kariyappa2021protecting}, NT~\cite{ma2021undistillable}, PoW~\cite{dziedzic2022increasing}, APGP~\cite{cheng2023apgp}, InI~\cite{guo2023isolation}) and three passive defenses (PRADA~\cite{juuti2019prada}, SEAT~\cite{zhang2021seat}, FDINET~\cite{yao2023fdinet}).

\textbf{Evaluation Metrics.}
(1) Benign Accuracy (BAcc.): percentage of correctly classified benign test data by protected model; (2) Clone Accuracy (CAcc.): percentage of correctly classified benign test data on clone model; (3) Detection Accuracy (DAcc.): percentage of successfully detected malicious samples; (4) False Positive Rate (FPR): percentage of benign samples incorrectly classified as malicious.

\textbf{Implementation Details.}
When a user's cumulative query count reaches $M_{0}=50k$, Model-Guardian terminates service if malicious queries exceed $K_{0}=50\%$. Adversaries are given query budgets of 20M (CIFAR-10, ImageNet) and 10M (CIFAR-100). Each experiment is run five times, with the average result reported. More details are in the Appendix-C.

\begin{table*}[t]
    \centering
    \caption{\textbf{Detection performance (\%) of DFMS-Detector on various GAN-generated images.} The meanings of each metric are consistent with Table \ref{tab:Comparison with Popular Passive Defenses}, and the batch size is 50.}
    \label{tab:Detection on GAN-Generated Images}
    \vspace{-0.1cm}
    \resizebox{\textwidth}{!}{
    \begin{tabular}{lcccccccccccccccc}
        \toprule
        \multirow{2}{*}{\textbf{Method}} & \multicolumn{2}{c}{\textbf{ProGAN}} & \multicolumn{2}{c}{\textbf{StyleGAN}} & \multicolumn{2}{c}{\textbf{StyleGAN2}} & \multicolumn{2}{c}{\textbf{BigGAN}} & \multicolumn{2}{c}{\textbf{CycleGAN}} & \multicolumn{2}{c}{\textbf{StarGAN}} & \multicolumn{2}{c}{\textbf{GauGAN}} & \multicolumn{2}{c}{\textbf{Total Avg.}} \\
        \cmidrule(lr){2-3} \cmidrule(lr){4-5} \cmidrule(lr){6-7} \cmidrule(lr){8-9} \cmidrule(lr){10-11} \cmidrule(lr){12-13} \cmidrule(lr){14-15} \cmidrule(lr){16-17}
        & DAcc. & FPR & DAcc. & FPR & DAcc. & FPR & DAcc. & FPR & DAcc. & FPR & DAcc. & FPR & DAcc. & FPR & DAcc. & FPR \\
        \midrule
        PRADA & 0.00 & 0.00 & 0.00 & 0.00 & 0.00 & 0.00 & 0.00 & 0.00 & 0.00 & 0.00 & 0.00 & 0.00 & 0.00 & 0.00 & 0.00 & 0.00 \\
        SEAT & 9.00 & 3.00 & 12.00 & 4.00 & 10.00 & 4.00 & 18.00 & 3.00 & 13.00 & 5.00 & 21.00 & 2.00 & 24.00 & 5.00 & 15.29 & 3.71 \\
        FDINet & 91.00 & 0.00 & 89.00 & 0.00 & 86.00 & 0.00 & 93.00 & 2.00 & 89.00 & 4.00 & 95.00 & 0.00 & 92.00 & 1.00 & 90.71 & 1.00 \\
        Model- & \multirow{2}{*}{\textbf{100.00}} & \multirow{2}{*}{0.00} & \multirow{2}{*}{\textbf{97.00}} & \multirow{2}{*}{1.00} & \multirow{2}{*}{\textbf{96.00}} & \multirow{2}{*}{2.00} & \multirow{2}{*}{\textbf{99.00}} & \multirow{2}{*}{0.00} & \multirow{2}{*}{\textbf{99.00}} & \multirow{2}{*}{2.00} & \multirow{2}{*}{\textbf{100.00}} & \multirow{2}{*}{0.00} & \multirow{2}{*}{\textbf{98.00}} & \multirow{2}{*}{1.00} & \multirow{2}{*}{\textbf{98.43}} & \multirow{2}{*}{0.86} \\
        Guardian \\
        \bottomrule
    \end{tabular}
    }
\end{table*}

\begin{table*}[t]
    \centering
    \caption{\textbf{Detection performance (\%) of DFMS-Detector on various diffusion-generated images.} The meanings of each metric are consistent with Table \ref{tab:Comparison with Popular Passive Defenses}, and the batch size is 50.}
    \label{tab:Detection on Diffusion-Generated Images}
    \vspace{-0.1cm}
    \resizebox{\textwidth}{!}{
    \begin{tabular}{lcccccccccccccc}
        \toprule
        \multirow{2}{*}{\textbf{Method}} & \multicolumn{2}{c}{\textbf{ADM}} & \multicolumn{2}{c}{\textbf{SD-v1}} & \multicolumn{2}{c}{\textbf{Guided}} & \multicolumn{2}{c}{\textbf{DALL-E}} & \multicolumn{2}{c}{\textbf{LDM}} & \multicolumn{2}{c}{\textbf{Glide}} & \multicolumn{2}{c}{\textbf{Total Avg.}} \\
        \cmidrule(lr){2-3} \cmidrule(lr){4-5} \cmidrule(lr){6-7} \cmidrule(lr){8-9} \cmidrule(lr){10-11} \cmidrule(lr){12-13} \cmidrule(lr){14-15}
        & DAcc. & FPR & DAcc. & FPR & DAcc. & FPR & DAcc. & FPR & DAcc. & FPR & DAcc. & FPR & DAcc. & FPR \\
        \midrule
        PRADA & 0.00 & 0.00 & 0.00 & 0.00 & 0.00 & 0.00 & 0.00 & 0.00 & 0.00 & 0.00 & 0.00 & 0.00 & 0.00 & 0.00 \\
        SEAT & 5.00 & 3.00 & 6.00 & 3.00 & 8.00 & 2.00 & 3.00 & 0.00 & 3.00 & 1.00 & 5.00 & 0.00 & 5.00 & 1.50 \\
        FDINet & 82.00 & 2.00 & 80.00 & 1.00 & 86.00 & 1.00 & 79.00 & 1.00 & 75.00 & 1.00 & 84.00 & 0.00 & 81.00 & 1.00 \\
        Model-Guardian & \textbf{100.00} & 0.00 & \textbf{94.00} & 2.00 & \textbf{97.00} & 1.00 & \textbf{95.00} & 1.00 & \textbf{98.00} & 1.00 & \textbf{98.00} & 1.00 & \textbf{97.00} & 1.00 \\
        \bottomrule
    \end{tabular}
    }
\end{table*}

\vspace{-0.1cm}
\subsection{Defending against Mainstream Stealing Attacks}
\label{section5.2}
\vspace{-0.1cm}
\textbf{Comparison with Popular Active Defenses.}
Table \ref{tab:Comparison with Popular Active Defenses} compares our method with other active defenses against seven attacks on CIFAR-10, CIFAR-100, and ImageNet. Our approach provides strong defense, with clone model accuracies ranging from 10\%$\sim$15\%, 2\%$\sim$10\% and 2\%$\sim$15\% on these datasets, respectively.
Importantly, Model-Guardian does not affect the benign accuracy of normal user queries, thanks to the low false positive rate of the DFMS-Detector. Even with occasional false positives, the class probability relationships in the perturbed output remain unchanged, ensuring correct classification.

\textbf{Comparison with Popular Passive Defenses.}
We also compare detection performance on CIFAR-10 with three passive defenses.
Since the other methods rely on the feature distribution of continuous queries, we vary batch sizes (50 and 500) to assess malicious query behavior.
As shown in Table \ref{tab:Comparison with Popular Passive Defenses}, our method achieves over 97\% detection accuracy and under 2\% false positive rate, owing to the strong generalization of our detector, which independently evaluates each sample. 

\vspace{-0.1cm}
\subsection{Performance of DFMS-Detector on Malicious Queries from Various Generative Models}
\vspace{-0.1cm}

In Table \ref{tab:Detection on GAN-Generated Images} and Table \ref{tab:Detection on Diffusion-Generated Images}, our DFMS-Detector demonstrates superior detection performance on all images generated by various GANs and diffusion models, with average detection accuracies of 98.43\% and 97.00\%, respectively.


\begin{table}[t]
    \centering
    \caption{\textbf{Effectiveness (\%) of Deceptive Predictions.} The results indicate the accuracy of clone model obtained by attacker in two situations: no defense and with only DPreds.}
    \label{tab:Effectiveness of Deceptive Predictions}
    \vspace{-0.1cm}
    \resizebox{0.48\textwidth}{!}{
    \begin{tabular}{lcccccc}
        \toprule
        \textbf{Method} & \textbf{DaST} & \textbf{MAZE} & \textbf{DFME} & \textbf{DFMS} & \textbf{DisGUIDE} & \textbf{QUDA} \\
        \midrule
        No Defense & 35.18 & 45.60 & 88.10 & 91.24 & 94.02 & 86.94 \\
        w/ DPreds & \textbf{15.88} & \textbf{26.36} & \textbf{53.53} & \textbf{62.07} & \textbf{61.15} & \textbf{53.03} \\
        \textbf{Decline rate} & 19.30 & 19.24 & 34.57 & 29.17 & 32.87 & 33.91 \\
        \bottomrule
    \end{tabular}
    }
\end{table}

\begin{figure}[t]
\centering
\vspace{-0.2cm}
\includegraphics[width=0.4\textwidth]{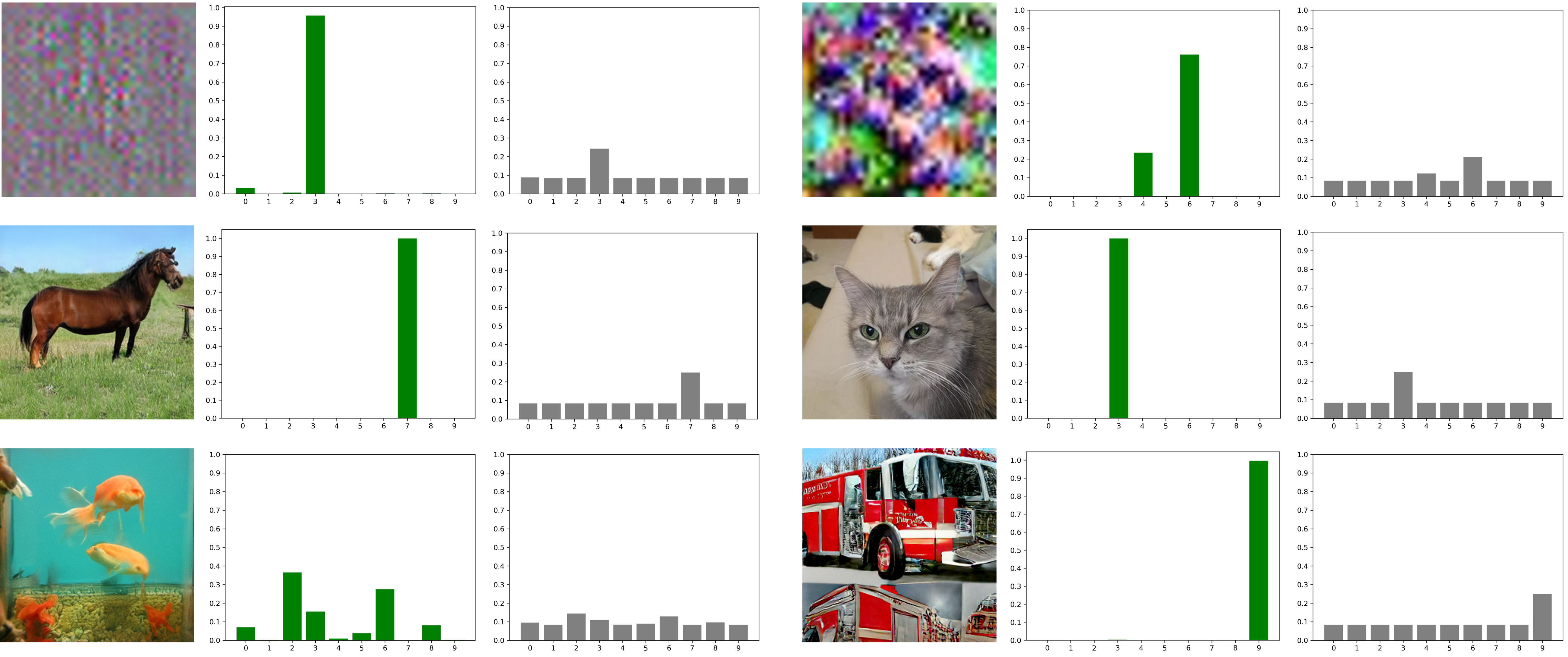}
\vspace{-0.1cm}
\caption{Visualization of class probabilities before perturbation (the second and fifth columns) and after perturbation (the third and last columns) on synthetic query images from six different sources (DaST, StyleGAN, ADM located in the first column and DFME, GauGAN, DALL-E located in the third column).}
\label{fig4}
\vspace{-0.3cm}
\end{figure}

\vspace{-0.1cm}
\subsection{Effectiveness of Deceptive Predictions}
\vspace{-0.1cm}
To evaluate DPreds' impact on model stealing defense, we test it independently against six data-free stealing attacks. As shown in Table \ref{tab:Effectiveness of Deceptive Predictions} with a perturbation value of 0.5, the module reduces clone model accuracy by 19.24\%$\sim$34.57\% across various attack methods.
For a clearer view of the perturbation algorithm's effect, we visualize the class probability distribution before and after perturbation in Figure \ref{fig4}. The visualization shows that while the class relationships remain intact, the probability values change, creating smoother peaks that mislead adversaries and lower clone model accuracy. Normal users remain unaffected, as they rely only on top-1 classification or the order of probabilities. Additionally, we provide t-SNE visualizations of the prediction distribution on CIFAR-10 before and after perturbations in the Appendix-D.

\vspace{-0.1cm}
\subsection{Other Experiments and Discussions}
\vspace{-0.1cm}
We also explore the impact of the transformation model, training data size, and perturbation value on our method, with results in Appendix-E. Discussions on hyperparameter selection, computational load, multi-adversary collusion, and extensibility are in Appendix-F.

\section{Conclusion}
\vspace{-0.1cm}
In this paper, we propose Model-Guardian, a defense against data-free model stealing attacks. By collecting synthetic and real data from three sources, we transform images into gradients, a more generalized representation. We then train binary classifiers and integrate them into the DFMS-Detector for malicious query detection. Additionally, we introduce the Deceptive Predictions algorithm to perturb outputs for malicious users. Extensive experiments show Model-Guardian outperforms existing defenses, effectively detecting covert malicious queries. Our work aims to inspire further research in defending against data-free model stealing attacks.


\bibliographystyle{IEEEbib}
\bibliography{icme2025references}


\appendix

\section{Appendix}

\begin{figure}[t]
\centering
\includegraphics[width=0.45\textwidth]{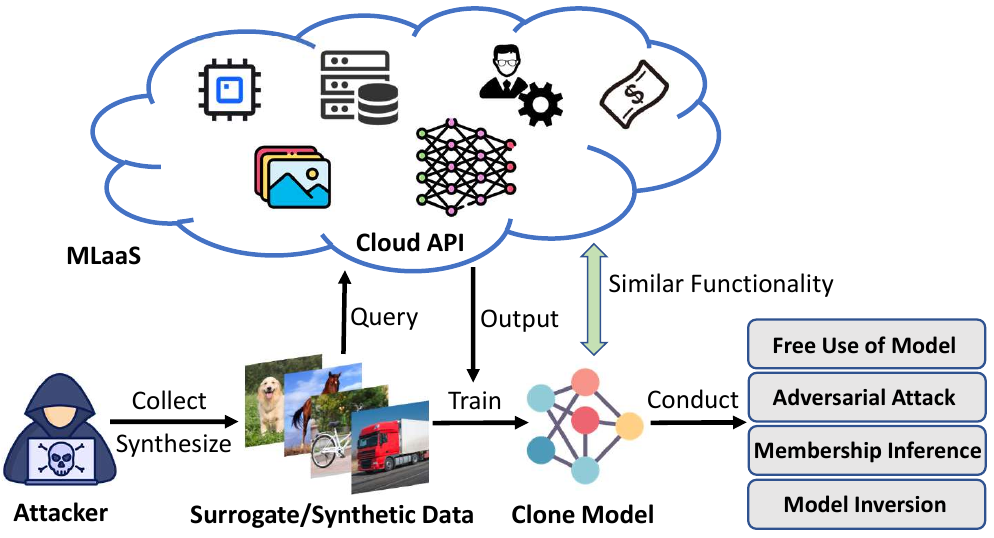}
\caption{\textbf{Model stealing attack and its vulnerabilities.} Attackers will query models deployed in the cloud through APIs using surrogate or synthetic data to obtain corresponding predictions. They can then use these annotated data to train a clone model with similar functionality to the original models and carry out further malicious actions.}
\label{fig1}
\end{figure}

\subsection{The Workflow of Model Stealing Attack}
As shown in Figure \ref{fig1}, model stealing attacks involve adversaries continuously querying APIs with crafted samples to acquire annotated data for training a functionally equivalent clone model. Beyond utilizing the stolen model without cost, attackers can employ it for adversarial attacks, membership inference attacks, and model inversion attacks.

\subsection{Insights on Defeating Data-Free Stealing}
\textbf{Fundamental Characteristic of Synthetic Images: Artifacts.}
In the context of Machine Learning as a Service (MLaaS), benign users query target models with natural, real data. However, a significant distinction of data-free model stealing attacks from previous attacks is the absence of real data, as they rely on synthetic data generated by generative models. Based on this fundamental difference, a straightforward and feasible defense against such attacks is to identify the synthetic queries used by attackers. Current data synthesis techniques, including generative adversarial networks (GANs) and diffusion models, produce data through generative processes. Both GANs and diffusion models involve numerous up-sampling operations in their workflows. Research in the field of deepfake detection \cite{zhang2019detecting,frank2020leveraging,durall2020watch,tan2024rethinking} has demonstrated that due to these up-sampling components, synthetic data generated by these models inherently carry unnatural features, known as artifacts. Artifacts refer to unnatural, artificially processed traces, or regions. Consequently, artifacts can serve as a reliable criterion for distinguishing between normal and malicious queries when defending against data-free stealing.

\textbf{Enhancing Defense Generalization: Transforming Data Dependency to Model Dependency.}
Building on the previous analysis, a straightforward method for detecting malicious samples involves collecting real and synthetic images to train a binary classifier as a detector. However, prior research \cite{wang2020cnn} has demonstrated that such detectors suffer from poor generalization, effectively identifying images generated by the same generator used during training but failing with images from other generators. This issue arises due to the detector's over-reliance on training data. To address this challenge, inspired by previous work \cite{wang2020cnn,li2022defending,liu2023detection,tan2023learning}, we propose using a more generalized representation, namely, the data gradients on a pre-trained model, as the detector's training data. Since these gradients retain only the essential pixels relevant to the pre-trained model's target task, they exclude the content of images. The remaining core pixels help the detector learn the common characteristics of this data type. Consequently, we transform the original data dependency problem into a model dependency problem. By converting synthetic data from different sources into gradient data using the same pre-trained model, we can reliably detect artifacts, ensuring the detector's effectiveness across various types of synthetic data.

\textbf{Defense vs. Utility: A Comprehensive Defense Approach.}
The aforementioned detector serves as just one component of our defense against data-free stealing attacks. Our ultimate objective is to prevent attackers from obtaining a well-performing clone model while ensuring that legitimate users can continue to use the protected model without disruption. To achieve this goal, we must consider multiple factors, including: (1) enhancing the reliability of detection results to minimize false positives, (2) managing malicious query samples and malicious users, and (3) maximizing user experience for benign users. Addressing these challenges necessitates the design of a comprehensive defense framework.

\subsection{Other Experimental Setup}

\begin{table}[t]
    \centering
    \footnotesize
    \caption{Composition of the training dataset for DFMS-Detector.}
    \label{tab:training set information}
    \begin{tabular}{lccc}
        \toprule
        \textbf{Dataset Name}   & \textbf{Real Source}   & \textbf{Fake Source} & \textbf{\makecell[c]{\# of images\\(real/fake)}} \\
        \midrule
        Data Subset 1  & TinyImageNet & DFME        & 20k/20k \\
        Data Subset 2  & LSUN          & ProGAN      & 20k/20k \\
        Data Subset 3  & ImageNet      & ADM         & 20k/20k \\
        \bottomrule
    \end{tabular}
\end{table}

\textbf{Composition of the Training Dataset for DFMS-Detector.}
In constructing the training set for DFMS-Detector, we collect a representative subset of samples from each of three different sources, which is a viable strategy for defenders. In detail, We simulate the data-free stealing process of DFME to query the defensive model, obtain a trained generator, and then input random noise to it to synthesize the synthetic samples of our first type of training data. The second type of synthetic training data is generated by a currently mainstream GAN model called ProGAN. The third type of synthetic data is synthesized using the popular diffusion model ADM. For each type of synthetic data, as samples with the class fake in the corresponding training subset, we additionally collect an equal amount of real samples with semantic class relevance to the synthetic data as samples with the class real in this training subset. The complete information regarding the training set is provided in Table \ref{tab:training set information}.

\textbf{More Implementation Details.}
During DFMS-Detector training, the Adam optimizer is employed with an initial learning rate of $5\times10^{-4}$. Decay involves multiplying the learning rate by 0.9 every ten epochs, with a total of 200 training epochs and a batch size of 32. Data augmentation methods, including random cropping, horizontal flipping, blurring, and compression, are utilized. The trained model serves as a classifier for the detector. 

\textbf{Experimental Platform.}
All our experiments are conducted on a server running the 64-bit Ubuntu 20.04.6 operating system. The server is equipped with an Intel(R) Xeon(R) Silver 4314 CPU @ 2.40GHz, 504GB of memory, and one NVIDIA A100 PCIe GPU with 40GB of memory. In the specific implementation of the source code, we use Python 3.10.4, Pytorch 1.11.0 and CUDA 11.8.

\begin{figure}[t]
\centering
\includegraphics[width=0.43\textwidth]{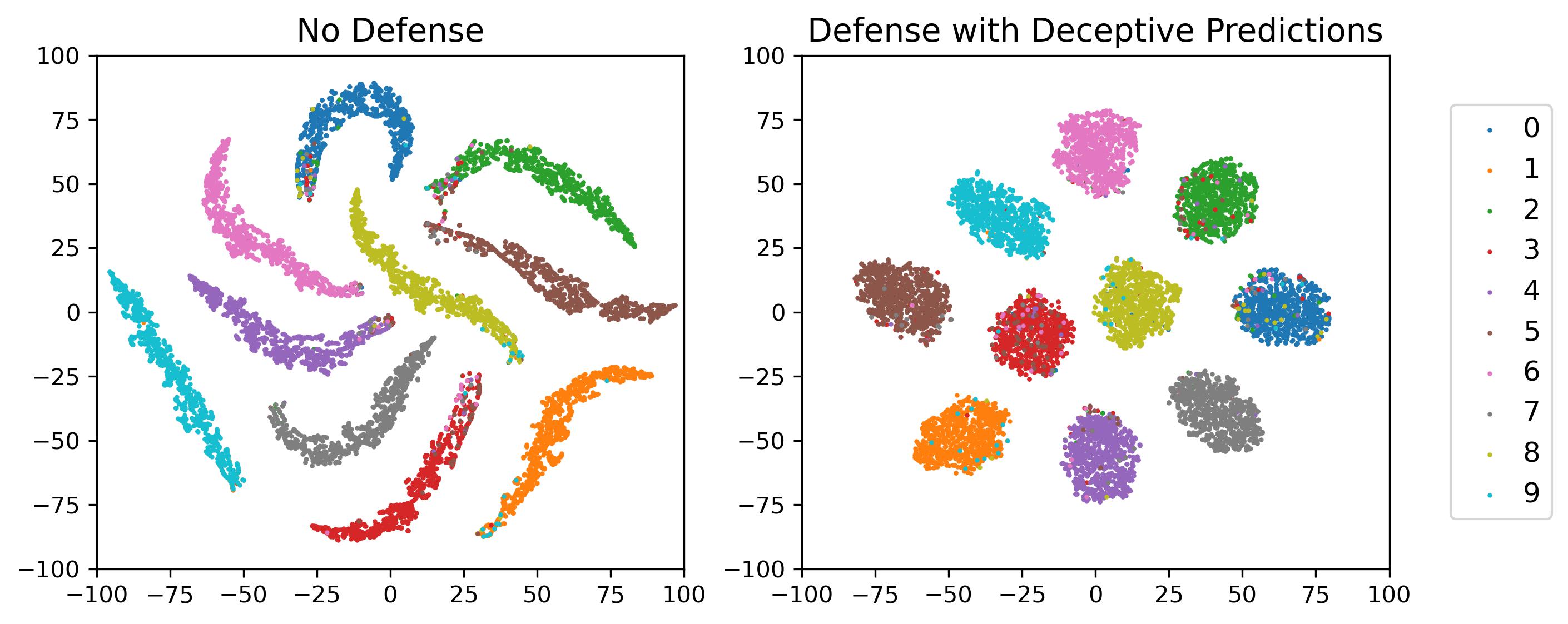}
\caption{Visualization of t-SNEs for both normal and defensive ResNet-34 on CIFAR-10. Each dot represents one data point.}
\label{fig5}
\end{figure}

\subsection{T-SNE Visualization for the Prediction Distribution of Victim Model on CIFAR-10}
We illustrate t-SNE visualization for the prediction distribution of the victim model on CIFAR-10 before and after perturbations in Figure \ref{fig5}. It is evident that after perturbations, although the model maintains normal classification, its output distribution undergoes a significant shift, deceiving adversaries' attempts to clone the victim model by mimicking its predictions.

\subsection{More Ablation Studies}
In this subsection, we conduct some ablation studies to further understand the functions of various modules of Model-Guardian, the impact of transformation models on the performance, the impact of the number of detector's training data on the performance, and the effectiveness of perturbation algorithm under different perturbation values.

\begin{table}[t]
    \centering
    \caption{\textbf{Ablation study (\%) on the contributions of different modules.} The results indicate the accuracy of clone model obtained by attacker using various data-free model stealing attacks under different defense settings.}
    \label{tab:Contribution of Different Modules}
    \resizebox{0.48\textwidth}{!}{
    \begin{tabular}{lccccc}
        \toprule
        \textbf{Method} & \textbf{MAZE} & \textbf{DFME} & \textbf{DFMS} & \textbf{DisGUIDE} & \textbf{QUDA} \\
        \midrule
        No Defense & 45.60 & 88.10 & 91.24 & 94.02 & 86.94 \\
        Model-Guardian & \textbf{11.27} & \textbf{11.81} & \textbf{12.95} & \textbf{14.93} & \textbf{13.72} \\
        w/o Gradient Trans. & 23.74 & 13.58 & 48.63 & 45.32 & 37.05 \\
        w/o DFMS-Detector & 26.36 & 53.53 & 62.07 & 61.15 & 53.03 \\
        w/o DPreds & 13.95 & 15.22 & 16.86 & 17.94 & 18.87 \\
        \bottomrule
    \end{tabular}
    }
\end{table}

\begin{table*}[ht]
    \centering
    \caption{\textbf{Detection performance (\%) of DFMS-Detector with different transformation models.} Since only the detection performance of our DFMS-Detector is evaluated here, the results in this table directly represent the overall detection accuracy and false positive rate on all test samples.}
    \label{tab:Detection Performance with Different Transformation Models}
    \resizebox{\textwidth}{!}{
    \begin{tabular}{lcccccccccccccccc}
        \toprule
        \multirow{2}{*}{\textbf{\makecell[c]{Transformation Model}}} & \multicolumn{2}{c}{\textbf{DaST}} & \multicolumn{2}{c}{\textbf{MAZE}} & \multicolumn{2}{c}{\textbf{DFME}} & \multicolumn{2}{c}{\textbf{DFMS}} & \multicolumn{2}{c}{\textbf{DisGUIDE}} & \multicolumn{2}{c}{\textbf{Dual Student}} & \multicolumn{2}{c}{\textbf{QUDA}} & \multicolumn{2}{c}{\textbf{Total Avg.}}\\
        \cmidrule(lr){2-3} \cmidrule(lr){4-5} \cmidrule(lr){6-7} \cmidrule(lr){8-9} \cmidrule(lr){10-11} \cmidrule(lr){12-13} \cmidrule(lr){14-15} \cmidrule(lr){16-17}
        & DAcc. & FPR & DAcc. & FPR & DAcc. & FPR & DAcc. & FPR & DAcc. & FPR & DAcc. & FPR & DAcc. & FPR & DAcc. & FPR \\
        \midrule
        Input Image & 64.82 & 8.34 & 79.18 & 7.12 & 97.43 & 4.59 & 61.09 & 6.56 & 76.34 & 9.59 & 71.95 & 9.78 & 56.92 & 9.88 & 72.53 & 7.98 \\
        ResNet18 & 73.06 & 7.28 & 85.22 & 4.66 & 96.38 & 3.06 & 69.28 & 4.63 & 83.12 & 6.84 & 80.02 & 7.79 & 65.13 & 6.43 & 78.89 & 5.81 \\
        CLIP-ResNet50 & 81.12 & 6.24 & 93.72 & 5.03 & 98.07 & 3.24 & 76.04 & 4.18 & 91.59 & 6.07 & 87.43 & 5.13 & 74.95 & 4.32 & 86.13 & 4.89 \\
        ProGAN-Discriminator & 83.35 & 7.16 & 95.46 & 6.05 & 98.12 & 2.90 & 85.77 & 3.23 & 94.03 & 5.34 & 92.75 & 5.10 & 78.62 & 4.19 & 89.73 & 4.85 \\
        StyleGAN-Discriminator & \textbf{91.47} & 6.05 & \textbf{97.16} & 5.23 & \textbf{98.36} & 2.32 & \textbf{93.75} & 3.09 & \textbf{96.37} & 5.01 & \textbf{95.58} & 3.65 & \textbf{91.74} & 3.52 & \textbf{94.92} & 4.12 \\
        \bottomrule
    \end{tabular}
    }
\end{table*}

\begin{table*}[ht]
    \centering
    \caption{\textbf{Detection performance (\%) of DFMS-Detector under different numbers of training data.} The meanings of each metric are consistent with Table \ref{tab:Detection Performance with Different Transformation Models}.}
    \label{tab:Detection Performance under Different Numbers of Training Data}
    \resizebox{\textwidth}{!}{
    \begin{tabular}{ccccccccccccccccc}
        \toprule
        \multirow{2}{*}{\textbf{\makecell[c]{Num. of\\Train. Data}}} & \multicolumn{2}{c}{\textbf{DaST}} & \multicolumn{2}{c}{\textbf{MAZE}} & \multicolumn{2}{c}{\textbf{DFME}} & \multicolumn{2}{c}{\textbf{DFMS}} & \multicolumn{2}{c}{\textbf{DisGUIDE}} & \multicolumn{2}{c}{\textbf{Dual Student}} & \multicolumn{2}{c}{\textbf{QUDA}} & \multicolumn{2}{c}{\textbf{Total Avg.}}\\
        \cmidrule(lr){2-3} \cmidrule(lr){4-5} \cmidrule(lr){6-7} \cmidrule(lr){8-9} \cmidrule(lr){10-11} \cmidrule(lr){12-13} \cmidrule(lr){14-15} \cmidrule(lr){16-17}
        & DAcc. & FPR & DAcc. & FPR & DAcc. & FPR & DAcc. & FPR & DAcc. & FPR & DAcc. & FPR & DAcc. & FPR & DAcc. & FPR \\
        \midrule
        5k & 85.97 & 8.13 & 91.17 & 9.92 & 90.92 & 8.35 & 90.38 & 7.56 & 92.33 & 9.97 & 89.49 & 6.93 & 85.22 & 8.49 & 89.35 & 8.48 \\
        10k & 87.28 & 7.40 & 94.32 & 8.68 & 94.66 & 5.77 & 92.02 & 4.87 & 93.05 & 9.43 & 92.83 & 7.45 & 89.35 & 6.06 & 91.93 & 7.09 \\
        20k & 88.92 & 6.56 & 96.05 & 6.02 & 95.97 & 3.38 & 91.80 & 5.64 & 95.14 & 6.72 & 95.26 & 5.24 & 91.08 & 5.43 & 93.46 & 5.57 \\
        40k & 91.47 & 6.05 & 97.16 & 5.23 & \textbf{98.36} & 2.32 & \textbf{93.75} & 3.09 & \textbf{96.37} & 5.01 & 95.58 & 3.65 & \textbf{91.74} & 3.52 & \textbf{94.92} & 4.12 \\
        60k & \textbf{91.83} & 6.22 & \textbf{97.21} & 5.39 & 98.19 & 2.54 & 93.14 & 3.71 & 95.84 & 5.32 & \textbf{96.02} & 3.51 & 91.59 & 4.12 & 94.83 & 4.40 \\
        \bottomrule
    \end{tabular}
    }
\end{table*}

\begin{table*}[ht]
    \centering
    \footnotesize
    \caption{\textbf{Active defense effects (\%) of DPreds under different perturbation values.} The results indicate the accuracy of clone model obtained by attacker using various data-free model stealing attacks under defense settings of different perturbation values.}
    \label{tab:Active Defense Effects under Different Perturbation Values}
    \begin{tabular}{cccccccc}
        \toprule
        \textbf{Perturbation Value} & \textbf{DaST} & \textbf{MAZE} & \textbf{DFME} & \textbf{DFMS} & \textbf{DisGUIDE} & \textbf{Dual Student} & \textbf{QUDA} \\
        \midrule
        0.00 & 35.18 & 45.60 & 88.10 & 91.24 & 94.02 & 91.36 & 86.94 \\
        0.05 & 19.04 & 24.41 & 71.15 & 75.73 & 79.93 & 74.82 & 70.44 \\
        0.10 & 16.26 & 22.90 & 72.34 & 74.52 & 77.04 & 71.90 & 68.56 \\
        0.20 & \textbf{14.71} & \textbf{21.23} & 66.04 & 70.59 & 72.52 & 67.35 & 65.21 \\
        0.50 & 15.88 & 26.36 & \textbf{53.53} & 62.07 & 61.15 & 57.56 & \textbf{53.03} \\
        1.00 & 15.24 & 26.56 & 56.15 & 59.25 & \textbf{60.58} & 56.83 & 53.26 \\
        2.00 & 14.95 & 23.35 & 53.60 & \textbf{58.42} & 61.87 & \textbf{56.04} & 54.61 \\
        \bottomrule
    \end{tabular}
\end{table*}

\textbf{Contributions of Different Modules.}
Table \ref{tab:Contribution of Different Modules} demonstrates that training the detector directly with images, without gradient transformation, results in reduced defense efficacy compared to the complete Model-Guardian. Particularly noteworthy is the significant improvement in effectiveness when DPreds module is employed for defense alone, compared to a scenario with no defense, underscoring the efficacy of the perturbation algorithm. Additionally, excluding the perturbation algorithm while retaining only the DFMS-Detector leads to a minor decrease in defense performance. This highlights the high detection accuracy of our detector, capable of promptly terminating adversary access to the service.


\textbf{Detection Performance of DFMS-Detector with Different Transformation Models.}
The transformation model plays an important role in implementing gradient transformation in our method, and here we explore the impact of its selection on the performance of the DFMS-Detector. Specifically, we select different types of models such as classification model, contrastive learning model, and discriminator of GAN as transformation models. We can draw three intuitive conclusions from the comparison results in Table \ref{tab:Detection Performance with Different Transformation Models}: (a) converting the training data of the detector from the input image to gradient data can improve the generalization ability of the detector, because gradient transformation filters out complex image content while retaining distinguishable key pixels; (b) The performance of GAN's discriminator is better than that of classification model and contrastive learning model, which reflects the natural advantage of GAN's discriminator in identifying artifacts in synthesized images caused by the up-sampling operation of the generative model; (c) The effectiveness of discriminators with different structures may also vary. These conclusions indicate that gradient transformation can transform the original data-dependency problem into a model-dependency problem, greatly improving the generalization of the DFMS-Detector. Given the best performance of the discriminator of StyleGAN, we will apply it as the final transformation model to our framework.

\textbf{Detection Performance of DFMS-Detector under Different Numbers of Training Data.}
In order to test the impact of different numbers of training data on the performance of the proposed DFMS-Detector, we collect training data of 5k, 10k, 20k, 40k, and 60k, with each setting having half the data number of the synthetic and real samples. During testing, we use 10k synthetic data and 10k real data each. Table \ref{tab:Detection Performance under Different Numbers of Training Data} demonstrates that as the number of training data increases, the detection performance of DFMS-Detector is better, but when it reaches above 40k, the performance increases very little and overfitting occurs. Therefore, the training data number of the DFMS-Detector we used in other experiments is fixed at 40k.

\textbf{Active Defense Effects of DPreds under Different Perturbation Values.}
We also explore the impact of different perturbation values on the effectiveness of our active defense method DPreds. In order to facilitate direct observation and obtain experimental results, we remove the DFMS-Detector and only retain the active defense module DPreds. From Table \ref{tab:Active Defense Effects under Different Perturbation Values}, it can be seen that as the perturbation value increases, the defense effect will gradually strengthen. In most attack methods, the defense effect reaches its peak when the perturbation value reaches 0.20 or 0.50. Even if the perturbation value continues to increase, the corresponding defense effect does not show a significant increasing trend, but gradually stabilizes.

\subsection{Discussion}
\textbf{Selection of Thresholds $M_{0}$ and $K_{0}$.}
$M_{0}$ and $K_{0}$ are critical hyperparameters in our approach, significantly influencing the overall performance of Model-Guardian. A larger $M_{0}$ means that the user's query count needs to accumulate to a higher number before deciding whether to terminate the service, potentially allowing the attacker to obtain a more accurate clone model. Conversely, a smaller $M_{0}$ can terminate malicious queries promptly but increases the likelihood of falsely terminating services for benign users. Similarly, a larger $K_{0}$ means more samples must be deemed malicious before service termination, reducing the detector's sensitivity and allowing attackers to evade detection. A smaller $K_{0}$ increases the detector's sensitivity but also raises the chance of false positives for benign users. Empirically, we set $M_{0}$ to $50k$ and $K_{0}$ to 50\% in our experiments, which we believe to be a reasonable combination, balancing the protection against attackers and the normal usage for benign users.

\textbf{Computational Cost.}
Our method requires each input sample to be transformed into a gradient image and uses a perturbation algorithm to modify the predictions of samples deemed malicious, introducing some computational overhead. We conduct a precise analysis of this computational cost. Specifically, converting a high-resolution $256\times256$ image to a gradient image takes only $7.70\times10^{-2}$ seconds, and perturbing a 10-class probability vector takes just $1.65\times10^{-5}$ seconds. These results indicate that the additional overhead introduced by our defense method is minimal, thus not limiting its practicality.

\textbf{Collusion Attacks with Multiple Accounts.}
To evade our detection, attackers might resort to registering multiple accounts or collaborating with others to conduct a collusion attack, where multiple accounts or attackers query the victim model to train a single clone model. Our method's inability to resist such collusion attacks is a limitation. However, collusion attacks require higher costs compared to ordinary attacks, and our DPreds module will continue to hinder the training of the clone model to some extent in this scenario, thereby reducing the effectiveness of collusion attacks.

\textbf{Flexibility and Extensibility.}
In this paper, we focus only on the image classification task. Future research could explore extending our method to defend against data-free model stealing attacks in other types of tasks. Given our framework's flexibility and ease of recombination, it can be adapted with minor modifications to apply to tasks such as object detection, semantic segmentation, image processing, natural language processing, and speech recognition. Additionally, replacing the DFMS-Detector component in Model-Guardian with other suitable anomaly detection algorithms can effectively defend against other types of model stealing attacks based on adversarial data or surrogate data. Overall, our method offers considerable flexibility and extensibility.

\end{document}